\documentclass{svmult}

\usepackage[english]{babel}
\usepackage[utf8x]{inputenc}
\usepackage{amsmath,amssymb}
\usepackage[comma,square,authoryear]{natbib}
\usepackage{graphicx}

\title*{Superprocesses as models for information dissemination in the Future Internet}

\author{Laura Sacerdote, Michele Garetto, Federico Polito, Matteo Sereno}
\authorrunning{L.\ Sacerdote et al.}
\institute{Laura Sacerdote (corresponding author) \at Dipartimento di Matematica, Universit\`a di Torino, Italy. \email{laura.sacerdote@unito.it} \smallskip \\
Michele Garetto \at Dipartimento di Informatica, Universit\`a di Torino, Italy. \email{michele.garetto@unito.it} \smallskip \\
Federico Polito \at Dipartimento di Matematica, Universit\`a di Torino, Italy. \email{federico.polito@unito.it} \smallskip \\
Matteo Sereno \at  Dipartimento di Informatica, Universit\`a di Torino, Italy. \email{matteo@di.unito.it}}

\begin{document}

\maketitle

\begin{abstract}\quad
    Future Internet will be composed by a tremendous number of potentially interconnected people and
    devices, offering a variety of services, applications and communication opportunities. 
    In particular, short-range wireless communications, which are available on almost all portable devices,
    will enable the formation of the largest cloud of interconnected, smart computing devices mankind has ever 
    dreamed about: the Proximate Internet. 
    In this paper, we consider superprocesses, more specifically super Brownian motion, as a suitable
	mathematical model to analyse a basic problem of information dissemination arising in the 
	context of Proximate Internet. The proposed model provides a promising analytical framework to both 
	study theoretical properties related to the information dissemination process and to devise efficient 
	and reliable simulation schemes for very large systems.	
\end{abstract}

\section{Introduction }

So many events had a significant impact on our life in these last 50 years
that it can be difficult to narrow it down to a few. Flights have
restricted our perception of distances, radio and television have made
people aware of facts and news, dangerous illnesses have disappeared thanks
to vaccines while new antibiotics have helped the recovery from many dangerous
diseases, medicine has decreased childhood mortality in many countries.
The list could continue but we focus here on one of the most sudden events
of our last thirty years that impacts our daily life everywhere: the advent
of the Internet. More than one billion people on Earth use the Internet nowadays.

Due to the importance of the Internet for our planet development, in Section
\ref{planet} we briefly list some important facts about it and its diffusion. Then in the
same section we present the past and the future role that wireless technology
played  and will play for the development of possible future services of the Internet.

Next, we move to the problem considered in this paper, i.e., the  
dissemination of information in large, disconnected mobile ad-hoc networks \citep{klein}. 
Such networks will arise naturally in a possible Future Internet scenario,
where an increasing number of users carrying existing and novel 
wireless devices (smartphones, tablets, laptops, smartwatches, smartglasses, etc.) 
will form the so-called \emph{Proximate Internet}, i.e., 
the largest cloud of interconnected, smart computing devices 
mankind has ever dreamed about.
In the \emph{Proximate Internet}, users will be able to directly communicate
among themselves exploiting short-range radio communications,
enabling a variety of interesting applications, and making 
the search and distribution of information much more efficient 
(in terms of spectrum usage, energy and monetary costs for the users)  
than traditional (cellular) networks.

In this scenario, information spreads among the wireless nodes 
in an epidemic fashion, i.e., like an infection in a human population.
Hence mathematical models developed for epidemiology 
can be applied, after properly adapting them to our specific context. 
Indeed, users carrying wireless devices are typically in motion
over a certain area (e.g., a city), and they can communicate with 
other devices within a limited communication range. Hence
the epidemic model must account for these fundamental features. 
 
The availability of mathematical models is of paramount importance 
to understand and design future applications running in the Proximate Internet. 
For example, models can predict the information dissemination speed, and thus 
the delays incurred to reach far-away users, and the 
efficiency of the dissemination in terms of area coverage, as function
of a variety of system parameters.      

The rest of the paper is organized as follows. In Section \ref{planet}
we start with a general introduction to the topic of mathematical 
modelling of the Internet. In Section \ref{diss1} we provide a high-level description 
of the specific scenario of wireless content distribution that we consider
in our work, before going into technical details. 
In  Section \ref{sbm}, we introduce some necessary background to the mathematical 
tools adopted in our model. In particular, we briefly recall how to
relate the discrete space and time branching process, i.e.\ the
Galton--Watson model, with its continuous time counterpart, i.e.\ the
continuous space branching process. Then we add to each
particle of the branching model a movement mechanism 
and we introduce the resulting super Brownian motion as a measure valued process
suitable to describe the information dissemination in a wireless cloud. 
For each of these processes we recall those properties which are most
significant for our modeling purposes.

In Section \ref{diss} we present our model as an instance of 
super Brownian motion, explaining why this representation provides
a much powerful performance evaluation methodology than 
traditional detailed simulations.
We conclude with 
directions of future research and a with a detailed reference section
which can be a helpful tool for the interested reader.

\section{Internet, Planet Earth and Mathematics}

\label{planet}
\begin{figure}
	\centering
	\includegraphics[scale=.5]{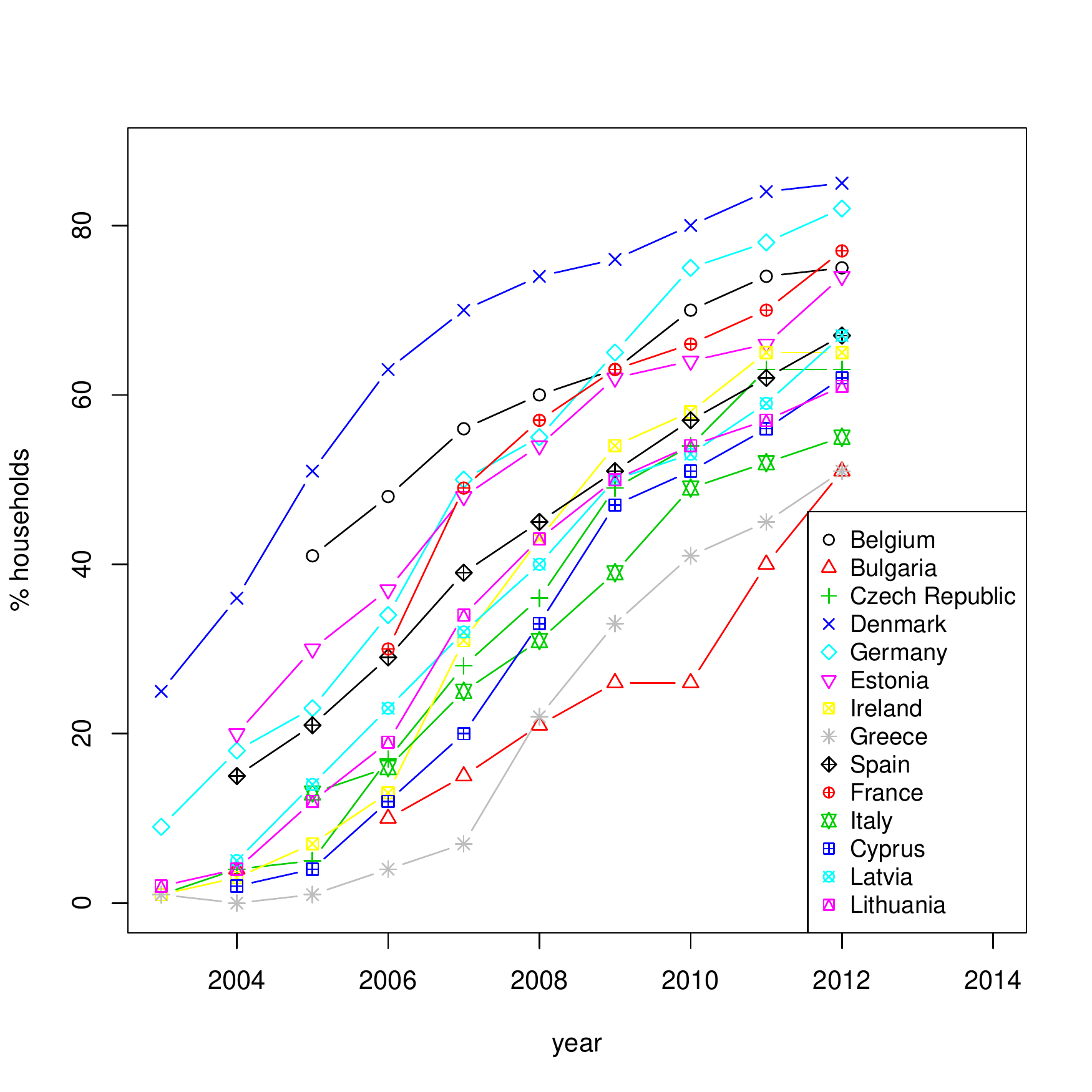}
    \caption{\label{grafico}Percentage of households with
    broadband access with at least one member aged 16 to 74 (source: Eurostat, see also Fig.\ \ref{grafica})}
\end{figure}

\begin{figure}
	\centering
   	\includegraphics[scale=.5]{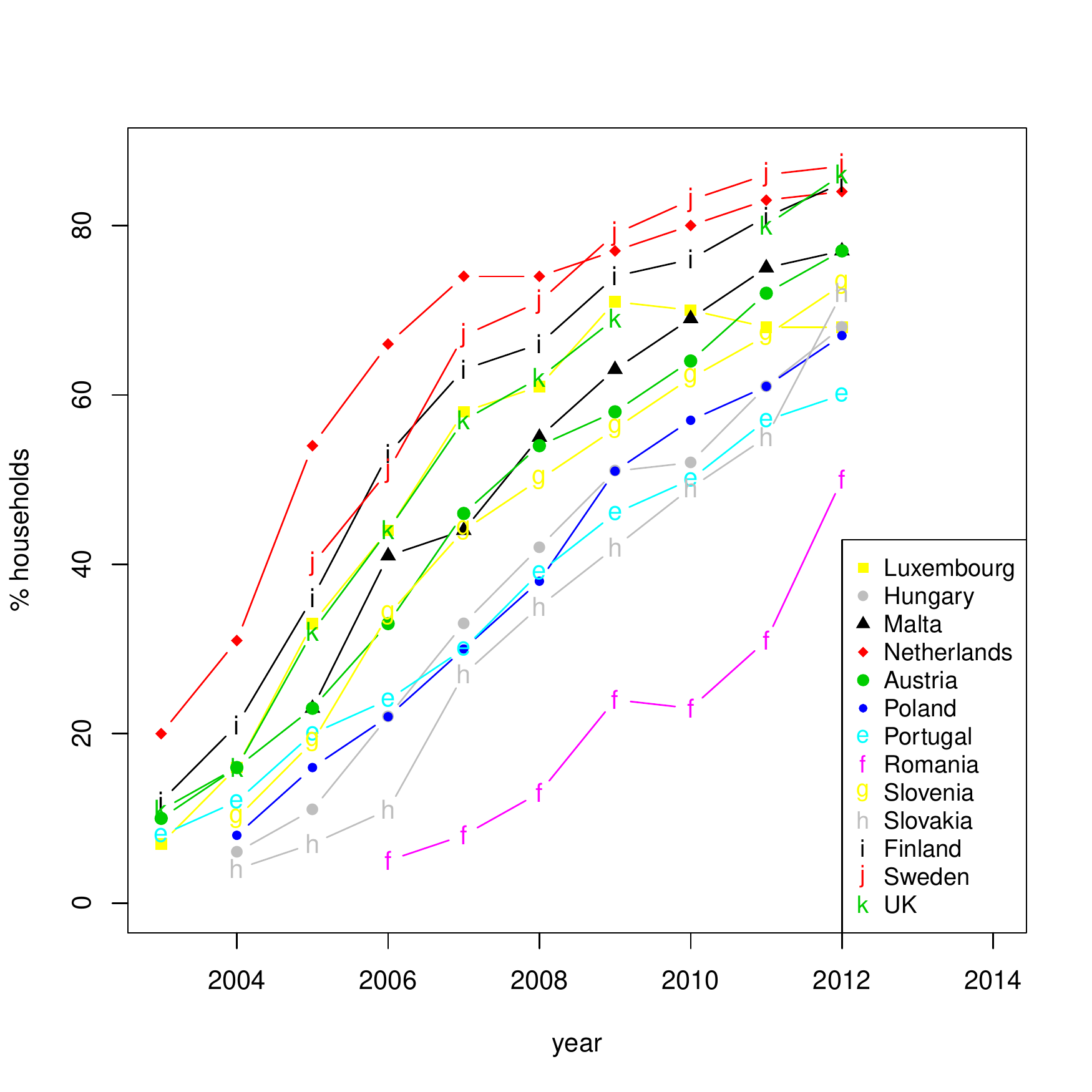}
    \caption{\label{grafica}Percentage of households with
    broadband access with at least one member aged 16 to 74 (source: Eurostat, see also Fig.\ \ref{grafico})}
\end{figure}

It is impressive to realize that the Internet has existed for a so short period
cosidering how strongly it changed the life on our Planet and the speed of
its penetration worldwide. In Fig.\ \ref{grafico} and \ref{grafica}
we illustrate the growth and the current presence of the Internet in Europe.
The Internet is also a formidable tool for the
evolution of underdeveloped countries where it allows the diffusion of books and
culture as well as to give long distance medical support to people living in
unreachable areas of our planet. The list of the changes determined in
everyday life is incredibly long and this short contribution is not the place
where to discuss such important political and social events.

We prefer to just provide an example taken from our everyday
experience as mathematical researchers, that clearly
show the fundamental role played by the Internet.   
The oldest between us still have memory of the incredible waste of time
related with any bibliographic research less then twenty years ago. Youngest
have never spent hours in a department library consulting books of
Mathematical Reviews, writing down the references and then moving to look for
the suggested journals, climbing ladders and moving heavy volumes up and
down. Now we simply Google the title or the subject of the paper
eventually using Mathematical Reviews online. When we recognize the title of
an interesting paper a further click miraculously let it appear on our screen. 
Going further, who has never checked the definition of some
mathematical object on Wikipedia for a fast suggestion?

The evolution of the Internet is so fast that we get immediately used to any
change and we finally disregard principia allowing the creation of these new
tools. However the improvements of the Internet are often related with
mathematical results since its first appearance. To cite a famous
example we recall that the first network between computers was made
possible by Kleinrock with his mathematical theory of packet networks, the
technology underlying today's Internet
\citep{kleinr}.
A more recent example is the mathematical work of \citet{brin} underlying the
algorithms of Google search engine (many mathematicians are still working on their improvements).

The Internet is also a source of new questions of high
mathematical interest. To give a couple of examples, let us think to the preferential attachment
phenomenon. In the Internet, preferential attachment makes nodes (that can 
represent either routers, or web pages, or people in social networks)
with higher degree to have stronger ability to grab links added to the network. 
It is well known that preferential attachment generates scale-free networks, i.e.\ networks
whose degree distribution follows a power law, at least asymptotically
\citep{barabasi,barabasi2, barabasi3}.
Various analytical results are known for this model but, for example, its
clustering coefficient is only numerically known and its analytical
expression is still an open problem. 

Many other examples of open mathematical problems could be cited, 
determined for example by congestion problems and transmission 
optimization \citep{srikant}.

We can pursue different philisophiae to deal with fundamental problems
related to the current and future Internet which translates into
different needs for mathematical models and
analyses. Next generation Internet involves different topics
that can be roughly classified under the primary networking function
to which they are related: routing, management and control (centralized or scalable),
security (additional overlap or part of the architecture), challenged
network environments (continuously connected--intermittent connectivity;
tools for connectivity), content delivery (robust and scalable methods) are
a possible classification of the involved topics.

Furthermore, each of these subjects can be studied both under the so called 
\emph{clean-slate} paradigm or as corrections and improvements of present situation. 
The \emph{clean-slate} design is the philosophy
adopted by a project started at Stanford University\footnote{Clean Slate Program, Stanford University, \texttt{http://cleanslate.stanford.edu}.}, based on the belief that the current
Internet has significant deficiencies that need to be solved 
before trying to improve future features of our global communication infrastructure.
The basic idea is to explore what kind of Internet we would design if we were to start
from scratch, with 20--30 years of hindsight. In this framework many interesting mathematical problems
arise. A good example coming from the area of network security 
are the so called \emph{adaptive resilient hosts}, based on the general idea that
a system should adapt to attacks by changing process code, perhaps also limiting
services temporarily\footnote{CRASH---Clean-slate Resilient Adaptive Secure Hosts, Cornell University, \texttt{http://www.nuprl.org/crash/introextended.php}.}. Other typical 
examples arise in the context of congestion control algorithms \citep{aliza}. 
On the other hand, many studies renounce the clean-slate approach, and just 
consider improvements of existing protocols and codes. 

Almost invariably, while modelling the Internet, evident difficulties arise 
both technically and mathematically, when the focus moves from an isolated 
problem to the whole Internet, which is 
an incredibly complex system whose evolution is not simply 
driven by technological factors. 

In our work we focus on a specific problem related to the transmission
of files to large populations of users. Hence we recall some basic 
facts about this topic. Wireless and wireline digital infrastructure currently coexist to transmit
digital and analog files of data, voice, video, text, image, fax and
streaming media. This coexistence will grow up in the next future giving rise
to new Internet services, day after day more user-friendly. Due to the growing
importance of the wireless technology, we provide here 
a brief summary of its history.
Wireless transmission has its origins around 1895 when
Tesla and Marconi independently developed the wireless telegraphy but its
mathematical foundations date back to 1864 when J.C.\ Maxwell mathematically
predicted the existence of radio waves. Mathematical results were then
experimentally verified by D.E.\ Hughes who first transmitted Morse code.
After wireless telegraph in 1900 wireless technology was used to transmit
voice over the first radio and following developments determined the
diffusion of radios around the globe.
Then wireless technology did not give rise
to new epochal instruments until the digital age of the Internet in the
latter part of the twentieth century, with the appearance of digital cellular
phones, mobile networks, wireless network access and so on. However the
beginning of the Internet was wireline based and only recently the power of
wireless communications has been unleashed.

L. Kleinrock in his paper on the History of Internet and its Flexible Future
\citep{kleinrock} recognizes the following five phases through which the
Internet will evolve in the next few years: nomadic computing, smart spaces and
smart networks, ubiquitous computing, platform convergence and intelligent
agents. Indeed there is an increasing number of travelling users that
requests trouble free Internet services from any device, any place and any
time (nomadic computing). Meanwhile it is no more science fiction to think to
a cyberspace where intelligent sensors give the alarm in the presence of
sudden health problems of aged people or where we can use voice commands to
interact with devices in a room (smart spaces and smart networks).
Ubiquitous computing is related with the necessity to have Internet services
available wherever, for example to consult maps or train timetables. As
Kleinrock remarks it is ridiculous to travel with a number of electronic
devices: cellular phones, notebook computers, clocks, microphones, cameras,
batteries, chargers. All these devices should converge in one, equipped with all
necessary technologies. Finally, intelligent agents,
i.e., autonomous software modules acting on data, observing trends,
carrying out tasks or adapting to the environment, will generate an
increasing amount of traffic that should use wireless channels.

Nowadays wireless technology is becoming absolutely fundamental 
both to access the Internet in the traditional sense and to 
communicate locally with nearby users and objects, 
like sensors and actuators (i.e., the Internet of Things). 
This evolution is made possible by the increasing popularity of 
portable devices, and the availability of a tremendous number of different services and 
applications enabled by these devices. 

\section{Wireless information dissemination: system overview} \label{diss1}
In our problem, we exploit device-to-device (wireless) communications 
to distribute information from one or more sources to a large number 
of mobile users belonging to the same \emph{cloud} (the Proximate Internet).
The exploitation of local, short-range communications (using for example WiFi or
Bluetooth technology) permits to offload cellular networks (3G/4G), with 
significant advantages both for cellular operators and for the users,
who can obtain almost for free the contents they are interested in 
directly from other nearby users.

More specifically, let us focus on a piece of information (an entire file or 
just a portion of it), hereinafter called the message, which is initially stored
at one or more nodes (called the seeds). These source nodes might have
retrieved the message from the Internet by other means (i.e., different wired or 
wireless technologies) or they can have generated it 
locally. The message can be transmitted to other nodes interested in it, but only
within a limited communication range, depending on the specific wireless 
technology (in the order or a few tens of meters). By so doing, the message
gets replicated in the wireless cloud in an epidemic fashion, reaching more and more
nodes with the passing of time. At the same time, nodes who have already received
the message may stop contributing to the dissemination process, due to 
several reasons (the application is closed, the wireless interface is 
switched off, the message is cancelled from the local memory to make room to 
other messages, etc). 
Furthermore, users can move while carrying their devices, and we
assume that the movements are random and independent from user to user.

Our goal is to define a suitable mathematical framework to model the above message
replication process in the case of a very large number of users, so that we can study 
the dynamics of wireless information dissemination over a large area (like a big city).
The model should be able to estimate basic performance metrics such as:

\begin{enumerate}
\item the delay associated to the transfer of the information, as function of the
distance from the source;

\item an index of the achieved city-level coverage, and in particular
a measure of the possible zones that will never be reached by the
information;

\item the probability that an uncovered zone will be covered again after a
reasonable delay.
\end{enumerate}

The availability of such a model would be very useful to plan and design
applications exploiting the proximate Internet. For example, it could be 
used to optimize the number and locations of the initial seeds,
trading off the various costs associated to bandwidth/energy/memory
constraints, according to different objective functions.   

\section{Mathematical background: Super Brownian motion}

	\label{sbm}
	In this section we present a brief description of the
    simplest superprocess that can be defined, arising as
    a weak limit of standard branching Brownian motions,
    the so-called super Brownian motion. In the most basic model
    of super Brownian motion we have critical behaviour, branching is quadratic, and jumps are not
    admitted. Plainly, more general models of super Brownian motions
    can be constructed, as we will see in the following. We start by analyzing the
    underlying branching structure and then we will add movement
    to the model.

\begin{figure}
    	\centering
        \includegraphics[scale=0.11]{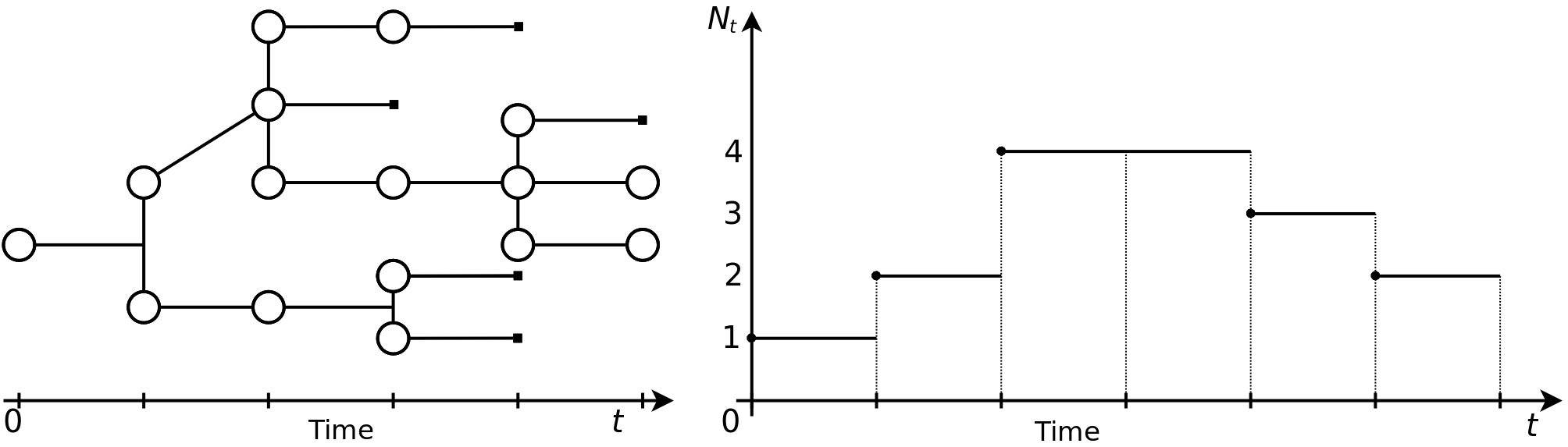}
    	\caption{\label{figura}
        	A realization of a Galton--Watson counting process
        	(right) with its associated genealogy tree (left).
            Births are indicated with circles
            while deaths are represented by black squares.
            Here time is continuous but $m = [t]$.
            In this figure it is possible to appreciate
            how a complete tree is far more rich in
            information than its related counting process.}
    \end{figure}

    Let us therefore first consider separately the branching structure.
    In practice the evolution of the number of nodes (i.e., the amount of messages disseminated in the network) 
    is governed by continuous state branching processes (CSBPs), introduced by \citet{jirina}
    and studied by various researchers during the past decades
    \citep[see for example][]{lamperti,lamperti2,feller,
    grey,silverstein,bingham,legall,li1,kale,liy,caballero,li2,
    kashikar}.
    In order to understand what a CSBP is, we start by describing a
    related discrete
    time and discrete state space
    model: the so-called Galton--Watson process.
	In general a Galton--Watson process
    \citep{watson,ney,jagers,haccou} models the
    evolution of a population composed by individuals acting independently
    and behaving homogeneously. Individuals live for only one generation
    (one time step) then they die producing a random number
    $Z$ of offsprings
    with a non negative discrete probability distribution
    $\mathbb{P}(Z=z)$, $z \in \mathbb{N} \cup \{ 0 \}$
    and with mean value $\mathbb{E}Z=\zeta < \infty$.
    Let us now write $N_m$, $m \in \mathbb{N}\cup \{ 0 \}$
    for the process counting the number of individuals in the
    population at time $m$. We put
    $N_0=1$ for simplicity (a single initial progenitor)
    and write that
    \begin{align}
    	N_{m+1} = \sum_{j=1}^{N_m} Z_j, \qquad m \in \mathbb{N} \cup \{0\},
    \end{align}
    where $Z_j$, are independent and identically distributed
    random variables representing the random number of
    offsprings generated by individual $j$.
    By means of
    simple calculations it
    can be proven that the mean value $\mathbb{E}N_m = \zeta^m$.
    The Galton--Watson process is critical, subcritical or
    supercritical if $\zeta=1$, $\zeta \in (0,1)$ or $\zeta > 1$, respectively.
    See Fig.\ \ref{figura} for an example of a realization
    of a Galton--Watson counting process together with the
    associated underlying tree coding the whole genealogy.
    For more information on Galton--Watson and related processes
    the reader can refer to the already cited classical o more
    modern references.

    Consider now a Markov process $Y_t^x$, $t \ge 0$,
    taking values in the positive real line,
    starting from point $x \in \mathbb{R} \cup \{ 0 \}$ and
    having right-continuous with left limits
    (c\`adl\`ag) paths.
    We refer to \citet{sharpe} or \citet{getoor} for an introduction to
    the general theory of Markov processes.
    We say that $Y_t^x$ satisfies the branching property
    if
    \begin{align}
    	Q_t(x+z,\cdot) = Q_t(x,\cdot) * Q_t(z,\cdot),
    \end{align}
    where $*$ represents the convolution operator and $(Q_t)_{t \ge 0}$ is
    the associated transition semigroup. This can be
    defined by its Laplace transform as
    \begin{align}
    	\label{lap}
    	\mathbb{E} \exp(-\mu Y_t^x) = \int_0^\infty \exp(-\mu y) \,
        Q_t(x,\mathrm dy) = \exp(-x \, v_t(\mu)), \qquad \mu > 0,
    \end{align}
    where $x \in \mathbb{R}^+ \cup \{ 0 \}$ is the starting point
    (initial population density) and the mapping
    $t \mapsto v_t(\mu)$ is the unique positive solution
    of the integral equation
    \begin{align}
    	v_t(\mu) = \mu - \int_0^t \phi (v_s(\mu)) \, \mathrm ds,
        \qquad t \ge 0.
    \end{align}
	The function $\phi$ in the previous equation is called the
    branching mechanism of the continuous state branching process
    and in general it can be of the form \citep{silverstein}
    \begin{align}
    	\label{barabas}
        \phi(z) = bz+cz^2 +\int_0^\infty (e^{-zy}-1+zy) \,
        \pi(\mathrm dy), \qquad z \ge 0,
    \end{align}
    where  $(y \wedge y^2)\pi(\mathrm dy)$ is a finite measure
    concentrated on the positive real line and $b \in \mathbb{R}$,
    $c \in \mathbb{R}^+ \cup \{ 0 \}$.
    Note that the above representation is related to the
    L\'evy--Khintchine representation for the characteristic function
    of a L\'evy process \citep{bertoin,sato,kyprianou}.
    The three different cases
    of $b>0$, $b<0$, and $b=0$ correspond respectively to the
    supercritical, subcritical, and critical cases.
    As we recalled above one of the simplest possible
    branching mechanisms is the quadratic (or binary) branching
    which corresponds to
    \begin{align}
    	\label{cdcd}
        \phi(z)=c z^2, \qquad c \in \mathbb{R}^+.
    \end{align}
    Continuous state branching processes with quadratic branching
    are of the critical type and are known in literature
    as the Feller's branching diffusions \citep{feller,pardoux}.
    From \eqref{lap} it is
    possible to determine the mean behaviour
    \begin{align}
    	\mathbb{E} Y_t^x = \int_0^\infty y \, Q_t(x, \mathrm dy)
        = x \, \exp(-bt), \qquad t \ge 0,\: x \in \mathbb{R}^+ \cup \{ 0 \},
    \end{align}
    for the general case \eqref{barabas}. When the branching
    mechanism reduces to \eqref{cdcd} we obviously obtain
    that $\mathbb{E}Y_t^x = x$ which characterizes a critical behaviour.
    In turn, when the branching is quadratic it is immediate
    to calculate an explicit form of the function
    $v_t(\mu)$ (see for example \citet{legall}) as
    \begin{align}
    	v_t(\mu) = \frac{\mu}{1+ct\mu}.
    \end{align}
    Continuous state branching processes are very well studied
    models of population evolution. For more in-depth
    results standard references include \citet{jirina},
    \citet{lamperti}, \citet{lamperti2},
    \citet{watanabe},
    \citet{silverstein}, \citet{bingham}, \citet[Chapter 3]{li},
    \citet{legall}, \citet{kyprianou}, and many others.
    Furthermore, it is worthy of notice that CSBPs arise naturally
    as weak limits of a sequences of rescaled Galton--Watson processes
    when waiting times between splits become negligible (see for example
    \citet{lamperti2}, \citet{haccou}, \citet{legall}).
    This last consideration will prove very useful
    in the interpretation of super Brownian motion.

    Consider therefore the $d$-dimensional Euclidean space $\mathbb{R}^d$
    and a random measure \citep[see][for reference]{kallenberg}
    on it determining the location of particles
    undergoing branching (both births and deaths) at time $t$, written as
    \begin{align}
    	\label{ques}
    	X_t = \frac{1}{\beta} \sum_j \delta_{B_t^j}, \qquad t \ge 0.
    \end{align}
    In \eqref{ques}, for each alive particle $j$, the process $B_t^j$,
    $t \ge 0$, is a $d$-dimensional Brownian motion on $\mathbb{R}^d$
    with initial starting point $x_0^j \equiv B_0^j$. Furthermore
    $B_t^j$ is independent of $B_t^i$, $i \ne j$, $\delta_{B_t^j}$
    is a Dirac delta measure on $B_t^j$, and $\beta$ is a normalizing
    factor related to the population size.
    The process \eqref{ques} is a measure valued  process on $\mathbb{R}^d$
    representing simultaneous motion of non-interacting particles.
    \begin{figure}
    	\centering
        \includegraphics[scale=0.6]{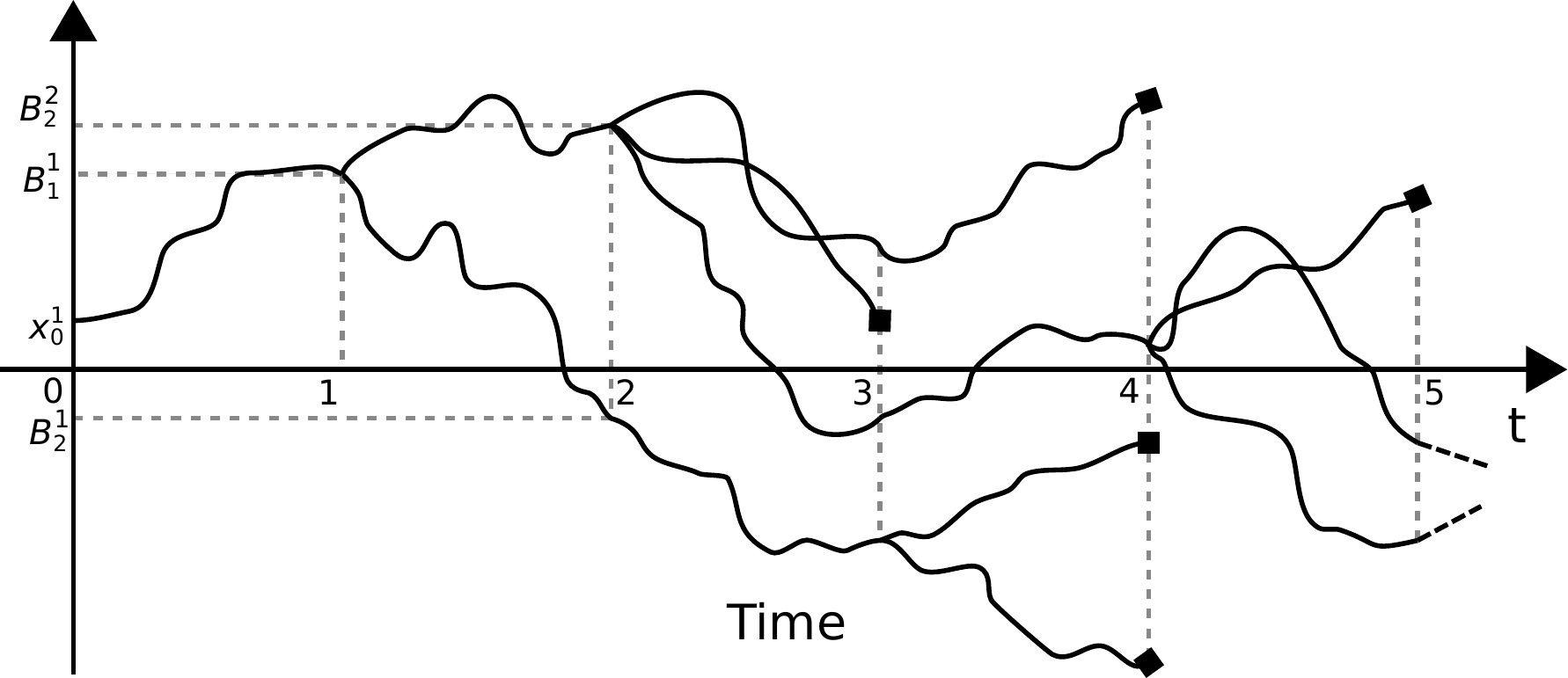}
    	\caption{\label{figura2}A sketch of a possible realization
        of the genealogy tree with Brownian movement. Death
        of particles is represented by black squares.}
    \end{figure}

    Beside this, each particle undergoes branching
    such that the total number of alive particles follows a
    Galton--Watson process:
    \begin{align}
    	W_t = \frac{1}{\beta}N_{[t]}, \qquad t \ge 0.
    \end{align}
    Note that branching occurs at time $1,2,3,\dots$, that is the
    waiting times between branching events are deterministic
    and all equal to unity.
    See in Fig.\ \ref{figura2} a sketch of a possible realization
    of the complete process where particles move following
    independent Brownian motions and branch at fixed times
    $t_1=1$, $t_2=2$, and so forth.

    We are in fact interested in the weak limit of the rescaled
    processes
    \begin{align}
    	X_t^k = \frac{1}{\beta_k} \sum_j \delta_{B_t^{j,k}}, \qquad
        t \ge 0, \: k \ge 1,
    \end{align}
    with $B_0^{j,k} \equiv x_0^{j,k}$.
    Clearly, aside the sequence $(X_t^k)_{k \ge 1}$, we consider
    the related sequence of processes
    \begin{align}
    	W_t^k = \frac{1}{\beta_k} N_{[kt]}^k, \qquad t \ge 0, \: k \ge 1,
    \end{align}
    which measures the total mass present (amount of messages in the network)
    and where the branching waiting times are all equal to $t/k$.
    The relation which links the two processes
    $W_t^k$ and $X_t^k$ is clearly that $\langle X_t^k,1 \rangle=W_t^k$.
    Let us now consider
    the space $M(\mathbb{R}^d)$ of finite measures on $\mathbb{R}^d$.
    If the sequence of the initial measures converges towards
    the finite measure $\xi \in M(\mathbb{R}^d)$ we have that
    $(X_t^k, t \ge 0)_{k \ge 1}$ converges weakly
    towards an $M(\mathbb{R}^d)$-valued Markov process $X_t$, $t \ge 0$
    (see \citet{lili}, Theorem 2.1.9, \citet{ethier}, Chapter 9).
    Correspondingly $(W_t^k, t \ge 0)_{k \ge 1}$
    converges weakly to
    a continuous state branching process with a general branching mechanism
    \eqref{barabas}.
    Since the underlying spatial motion is of Brownian type,
    the process obtained is called \emph{$\phi$-super Brownian motion}.
    In the simplest case of a quadratic (or binary) branching
    it is simply called \emph{super Brownian motion}.
    Relevant references for super Brownian motion and
    superprocesses in general are \citet{dawson}, \citet{li},
    \citet{legall}, \citet{etheridge}, \citet{perkins}, \citet{legall2},
    \citet{fitzsimmons}, \citet{kyp}, \citet{morters}.
	Moreover, it should be noted that also
    more general superprocess are already described in the literature
    \citep[see e.g.][]{englander,eng2,fleischmann,dawson2,wang}.

\section{Wireless information dissemination: basic model and future directions}
\label{diss}
Now we can explain how the information dissemination scenario 
introduced in Section \ref{diss1} can be represented in terms of a 
super Brownian motion of the simplest kind, as described in Section \ref{sbm}.

To do so, some simplifying hypotheses are necessary
in order to obtain a tractable model of the considered system.
We assume that each user independently moves over the infinite plane
according to a simple random walk. We know that realistic models 
of human mobility suggest that individuals' movements should be
modeled through L\'evy flights, i.e.\ as a random walk in which the
step-lengths have a probability distribution that is heavy-tailed \citep{brockmann}.
Alternatively truncated Levy flights were proposed after analysing data \citep{barabasi2}. 
However, for a first simplified model we consider the movements of each user  
as a simple random walk, allowing us to adopt a standard Brownian motion
to describe the macroscopic mobility of each node.  

While moving, nodes come in contact with other nodes (i.e., other nodes 
fall in the communication range of the considered node), allowing the opportunistic, 
direct transfer of the message of interest. A standard computation, that takes jointly into 
account the node density, the transmission range, and the node speed, 
permits computing the rate $\lambda$ at which the message gets 
duplicated. We do not repeat the details of this computation, which can be found in 
\citep{madhow}. Hence, the birth rate $\lambda$
can be considered as a primitive system parameter.

We instead denote by $\mu$ the death rate of each node (i.e., the rate at which a node transits
to a state in which it no longer contributes to the message dissemination), which
is supposed to be given.
At last, let $\sigma^2$ be the infinitesimal variance of the
Brownian motion describing the macroscopic mobility
of each node. 
Moreover, we assume that, at time $t=0$, there is a given 
number $N_0$ of seeds, initially co-located at the origin of the plane. 

Disregarding for now the nodes' movements, these hypotheses
characterize the previously presented \emph{Galton--Watson tree}, i.e.\ the model
that describes the evolution of a population that starts with a given size 
at time $t=0$. 
Depending on the value of $\lambda$ with respect to $\mu$ the population
becomes extinct with probability $1$ (subcritical case $\lambda<\mu$ and
critical case $\lambda=\mu$) or has a positive probability to survive
forever (supercritical case: $\lambda >\mu$). When the population becomes
sufficiently large its size can be modelled as a \emph{continuous state
branching process}, as we pointed out in the previous paragraph. Hence the
number of nodes storing the message of interest can be described
through these processes.

However, nodes also move according to a random walk, and we are
especially interested in characterizing the distribution on the plane of 
the nodes who are currently storing the message, as function of time.
In our brief review in Section \ref{sbm} we pointed out that, when the size of the
population is large and microscopic displacements of the nodes are frequent enough, 
one can rescale the Galton--Watson tree with moving particles obtaining a \emph{super Brownian
motion}.

For the super Brownian motion, results are available for the asymptotic speed
of diffusion. These results can be directly used to study the delay incurred by the
message delivery to far-away users.

In our case nodes move over $\mathbb{R}^{2}$ and the super Brownian motion 
is known to have singular distribution on $\mathbb{R}^{2}$.
Furthermore this distribution is uniform on its random support. Then
the study of the support properties of a super Brownian motion is a possible
approach to deal with the city coverage problem: we assume that each 
device covers a region of radius $r$, equal to its transmission range.
This leads to study the coverage of the support of the Super Brownian Motion 
through balls. The aim of this study should be to determine which percentage of the
plane is covered by at least one ball. Possible zones that will never be reached 
by the information could also be investigated following this approach.

Beside the above theoretical results, our model also allows us to
devise efficient simulation schemes for scenarios in which no analytical results 
are available, such as those in which the birth and/or death rates of the nodes,
and/or the infinitesimal variance of the associated brownian motion, are 
a function of space and/or time. 
On this regard, we emphasize that a brute-force simulation approach in which
each node is modelled in all details becomes infeasible for large number of nodes.
In particular, in the super-critical regime it is basically impossible
to simulate the system after a very short time, 
due to the exponential growth of the number of nodes.
Therefore, alternate simulation approaches are necessary.

Simulations represent the only methodology to study scenarios of 
particular interest, in which the parameters of the process are not homogeneous in space 
(presence of lakes, zones without inhabitants and so forth). The available
techniques for these simulations do not consider the singular nature of
the super Brownian motion. Hence reliable and efficient simulation methods 
will be the subject of future work, together with the study 
of more theoretical results of interest for our application scenario.

\subsubsection*{Aknowledgements}

The authors have been supported by project AMALFI \\
(Universit\`{a} di Torino/Compagnia di San Paolo).
 
\bibliographystyle{plainnat}
\bibliography{main}

\end{document}